\documentclass[aps,prl,twocolumn,amsmath,amssymb,nofootinbib,superscriptaddress]{revtex4}

\usepackage{color}

\newcommand{\bra}[1]{\langle#1|}
\newcommand{\ket}[1]{|#1\rangle}

\usepackage[pdftex]{graphicx}
\usepackage{mathrsfs}
\usepackage{times}

\begin{document}

\bibliographystyle{apsrev}

\title{Quantum walks with encrypted data}

\author{Peter P. Rohde}
\email[]{dr.rohde@gmail.com}
\homepage{http://www.peterrohde.org}
\affiliation{Centre for Engineered Quantum Systems, Department of Physics and Astronomy, Macquarie University, Sydney NSW 2113, 
Australia}

\author{Joseph F. Fitzsimons}
\affiliation{Centre for Quantum Technologies, National University of Singapore,
Block S15, 3 Science Drive 2, Singapore 117543}

\author{Alexei Gilchrist}
\affiliation{Centre for Engineered Quantum Systems, Department of Physics and Astronomy, Macquarie University, Sydney NSW 2113, 
Australia}

\date{\today}

\frenchspacing


\begin{abstract}
In the setting of networked computation, data security can be a significant concern. Here we consider the problem of allowing a server to remotely manipulate client supplied data, in such a way that both the information obtained by the client about the server's operation and the information obtained by the server about the client's data are significantly limited. We present a protocol for achieving such functionality in two closely related models of restricted quantum computation -- the Boson sampling and quantum walk models. Due to the limited technological requirements of the Boson scattering model, small scale implementations of this technique are feasible with present-day technology.
\end{abstract}

\maketitle


\emph{Introduction ---} Quantum information processing \cite{bib:NielsenChuang00} allows certain key problems, which are believed to be classically hard, to be efficiently solved. Well known examples with real world applications include Shor's algorithm for integer factorisation \cite{bib:Shor97} and Grover's search algorithm \cite{bib:Grover96}. One of the more promising approaches to implementing quantum algorithms is linear optics quantum computation (LOQC) \cite{bib:KLM01,bib:KokLovett11}, where information is encoded into single photons and the their wave properties are manipulated using linear optics elements. Photons are ideally suited to communication, leading naturally to models of distributed quantum computation.

A key consideration in any distributed computation scheme is security. Consider two parties, Alice and Bob. Alice has some data to which she would like to apply a computation, whilst Bob has a quantum computer and an algorithm with which he can process the data. However both sides have proprietary knowledge. Alice wants to keep her data secret from others, and Bob wants to keep his algorithm secret. This is related to the problem of \emph{homomorphic encryption} which allows data to be manipulated without decrypting, so Bob can perform a universal set of operations on Alice's data without ever learning Alice's input state. Universal classical homomorphic encryption was only first discovered in 2009 \cite{gentry2009fully} and subsequently simplified \cite{van2010fully}. Closely related is \emph{blind computing}, where Alice possesses both the data and the algorithm, and Bob owns the computer \cite{bib:blind2,bib:blind3,bib:blind1}, as is the quantum private queries protocol \cite{giovannetti10}, which is used to query a database while keeping the query secret.

In this paper we describe a technique for solving the above problem, and hence achieving a limited quantum homomorphic encryption using the Boson sampling and multi-walker quantum walk models for quantum computation.


\emph{The Boson sampling model ---} A first protocol for universal LOQC was introduced by Knill, Laflamme \& Milburn (KLM) \cite{bib:KLM01}. While universal for quantum computation, their protocol is extremely demanding, requiring fast-feedforward and quantum memory, which are technologically challenging and well beyond the capabilities of present-day experiments. Since then numerous simplifications have been proposed, most notably approaches based on cluster states \cite{bib:Raussendorf01,bib:Raussendorf03,bib:Nielsen04}, which significantly reduce physical resource requirements. However they remain very demanding to implement.

Recently Aaronson \& Arkhipov \cite{bib:AaronsonArkhipov10} introduced a much simplified model for LOQC, known as the Boson sampling model. While not believed to be universal, it was shown that this protocol very likely implements an algorithm which cannot be efficiently classically simulated (efficient classical simulation would likely imply a collapse in the polynomial hierarchy, \textbf{PH} \cite{bib:AaronsonArkhipov10}). The protocol does away with fast-feedforward and quantum memory, requiring only a multi-photon input state, a purely linear optics network, and photo-detection.

In the photon number basis, the input state is of the form \mbox{$\ket{\psi_\mathrm{in}} = \ket{1_1,\dots,1_{p},0_{p+1},\dots,0_{m}}$},
or any permutation thereof, where there are $p$ photons and $m$ modes. To the input state a unitary map is applied, which implements the transformation \mbox{$a_i^\dag \to \sum_j U_{ij} a_j^\dag$} on the photon creation operators. It was shown by Reck \emph{et al.} \cite{bib:Reck94} that any such $U$ can be efficiently constructed using a linear network comprising only beamsplitters and phase-shifters.

In an occupation number representation, the output state is of the form \mbox{$\ket{\psi_\mathrm{out}} = \sum_S \gamma_S \ket{n_1^{(S)},n_2^{(S)}\dots,n_N^{(S)}}$}, where $S$ are the different photon number configurations, $\gamma_S$ are the associated amplitudes, and $n_i^{(S)}$ is the number of photons in mode $i$ given configuration $S$. Each amplitude is proportional to a matrix permanent, whose calculation resides in the complexity class \textbf{\#P}-complete, giving rise to the believed classical hardness of calculating the output distribution.


\emph{The multi-walker quantum walk  model ---} Another interesting approach to LOQC is the quantum walk model \cite{bib:ADZ,bib:AAKV,bib:Kempe08}. Here our physical system comprises a graph in which walkers (i.e. photons) are placed at vertices and are allowed to coherently `hop' along the edges. The restriction to linear optics means that we consider only non-interacting walkers. The evolution is decomposed into two stages -- \emph{coin} ($C$) and \emph{step} ($S$) operations. The coin coherently manipulates an ancillary parameter known as the \emph{coin} value, while the step operator updates the \emph{position} (i.e. vertex) of the walker according to the direction specified by the coin. The evolution of the system proceeds by repeated application of coin and step, $\ket{\psi_\mathrm{out}}=(SC)^t\ket{\psi_\mathrm{in}}$. Rohde \emph{et al.} \cite{bib:RohdeSchreiber10} recently introduced a formalism for multi-walker quantum walks on general graphs. Indeed, numerous authors have begun experimentally demonstrating elementary optical quantum walks \cite{bib:Schreiber10,bib:Broome10,bib:Peruzzo10,bib:Schreiber11b,bib:Schreiber12}.

It can be shown that any unitary map on the photon creation operators can be decomposed into a non-interacting quantum walk, and similarly any non-interacting quantum walk can be expressed as such a unitary network \cite{bib:RohdeSchreiber12}. As with Boson sampling, no measurement or feedforward is performed within the evolution of the quantum walk. Thus there is a natural isomorphism between the two formalisms. We therefore refer to Boson sampling and multi-walker quantum walks on general graphs interchangeably. Boson sampling can be regarded as a classically hard task performed by a quantum walk.


\emph{Homomorphically encrypted Boson sampling and quantum walks ---} The first step in our protocol is to encode the Boson sampling input state into the polarisation basis. Suppose there are $m$ modes. Then for every mode in which a photon should be present we introduce a photon in the horizontal polarisation ($H$), and for every mode in which no photon should be present we introduce a photon in the vertical polarisation ($V$). Thus, there are always exactly $m$ photons in the system and the number of $H$s in the input state is equal to the number of photons in the corresponding non-polarisation-encoded state. For example, if the Boson sampling computer is supposed to be initialised with the input state $\ket{0,1,1,0,0,1}$, we would encode this using 6 photons as $\ket{\psi_\mathrm{in}}=\ket{V,H,H,V,V,H}$. Next we note that if we employ polarisation-resolving photo-detection at the output, and only measure those photons in the $H$ polarisation while discarding all $V$ photons, the operation of the circuit is identical to the desired Boson sampling computer, since $H$ and $V$ photons will not interfere. On the other hand, if we employ non-polarisation-resolving detectors, the output will effectively be corrupted.

Alice begins by preparing an encoded input state \mbox{$\ket{\psi_\mathrm{encoded}} = R(\frac{k\pi}{d})^{\otimes m} \ket{\psi_\mathrm{in}}$}, where \mbox{$R(\theta) = \left( \begin{array}{cc} \mathrm{cos}\,\theta & -\mathrm{sin}\,\theta \\ \mathrm{sin}\,\theta & \mathrm{cos}\,\theta \end{array}\right)$} is a polarisation rotation operator, which can be implemented using wave-plates, $d$ is the number of \emph{divisions} in the choice of rotation angle, and $k$ represents the $k$th division. Alice chooses $k$ randomly in the range $0$ to $d-1$. $k$ can be regarded as Alice's private key. Thus from Bob's perspective, the encoded state is a mixture of input states rotated by different angles, and it is this added noise that will allow Alice to hide her data from Bob. With $d$ divisions, the basis of each choice of encoded state is rotated by $\pi/d$ from the previous. The choice of $k$ is retained only by Alice, while the encoded state is communicated to Bob, who, not knowing the basis in which to measure, perceives a mixed state. At the end of the computation Alice measures the output state in the polarisation basis given by $R(\frac{k\pi}{d})$, allowing perfect reconstruction of the desired output state using polarisation-resolving photo-detection.


\emph{Information theoretic analysis ---} We now consider the security of our protocol in the context of Bob's probability of correctly inferring Alice's input state. To do so we calculate the Holevo information \cite{bib:NielsenChuang00} of the state sent from Alice to Bob. The Holevo quantity provides an upper bound on the amount of information Bob can extract from Alice's encoded state. Formally, the Holevo quantity of our protocol is given by
\begin{equation*}
\chi(m) = -\mathrm{Tr}(\rho \, \mathrm{log}_2\rho) + \frac{1}{2^{m}} \sum_{i=0}^{2^{m}-1} \mathrm{Tr}(\rho_i\, \mathrm{log}_2 \rho_i),
\end{equation*}
where $\rho = \frac{1}{2^{m}} \sum_{i=1}^{2^{m}} \rho_i$, and $\rho_i = \sum_{k=0}^{d-1} \bigotimes_{j=1}^{m} R\left(\frac{k\pi}{d}\right) \ket{P_{ij}}\bra{P_{ij}} R\left(-\frac{k\pi}{d}\right)$, 
and $\ket{P_{ij}}=\ket{H}$ when the $j^\mathrm{th}$ bit of $i$ is 0, otherwise $\ket{P_{ij}}=\ket{V}$.


While a closed form for the Holevo information for arbitrary values of $d$ and $m$ is likely too much to hope for, we can calculate the scaling of the Holevo information for $d \gg m$. To do this, we first note that since $\bigotimes_{j=1}^{m} \ket{P_{ij}}$ for the various values of $i$ form a complete basis on the space of input states, $\rho$ is the maximally mixed state. Therefore $-\mbox{Tr}(\rho \log_2 \rho) = m$. Next we note that $-\mbox{Tr}(\rho_i \log_2 \rho_i)$ is independent of $i$, and hence it is sufficient to consider only the case of $i=0$. We consider the change of basis $\ket{0} = (\ket{H} +i\ket{V})/\sqrt{2}$, $\ket{1} = (\ket{H} -i\ket{V})/\sqrt{2}$. As $\rho_0$ is a mixed state of symmetric states, it resides entirely in the symmetric subspace, which has dimension $n+1$. Thus a complete basis is formed by the states $\ket{\ell}_{m}$, the symmetric state of $m$ qubits containing exactly $\ell$ qubits in state $\ket{1}$, and the rest in state $\ket{0}$. In this basis, the density matrix $\rho_0$ is given by
\begin{equation*}
\rho_0 = \frac{1}{2^{m}} \sum_{k=0}^{d-1} \sum_{a,b = 0}^{m} e^{i\frac{(b-a)k\pi}{d}} \sqrt{\binom{m}{a}\binom{m}{b}} \ket{a}_{m}\bra{b}_{m}.
\end{equation*}
From this, we can see that the cross terms go to zero for large $d$ since in this case $\sum_{k=0}^{d-1} e^{i \frac{(b-a)k\pi}{d}} \to 0$. In such a case the density matrix is diagonal, and hence we have
\begin{equation*}
\mbox{Tr}\left( \rho_i \log_2 \rho_i \right) = \frac{1}{2^{m}} \sum_{a=0}^{m} \binom{m}{a} \log_2 \left(\frac{1}{2^{m}}\binom{m}{a}\right),
\end{equation*}
which is simply the entropy of the binomial distribution. This value is known to be $\frac{1}{2}\log_2\left(\frac{1}{2}\pi e m\right) + O(1/m)$, and hence the Holevo quantity scales as
\begin{equation*}
\chi(m) = m- \frac{1}{2}\log_2\left(\frac{1}{2}\pi e m\right) + O\left(\frac{1}{m}\right).
\end{equation*}
Hence the protocol hides $\frac{1}{2}\log_2\left(\frac{1}{2}\pi e m\right) + O(1/m)$ bits of information for suitably large $d$.

We note that if Bob has no prior information about Alice's chosen state, the probability that Bob correctly infers Alice's state can be bounded as follows. Let $\rho_X$ be the density matrix Bob receives from Alice when her input string is $X$. Bob must make a measurement on this state to determine his guess for $X$, which we denote $\tilde{X}$.
Without loss of generality we can view Bob's measurement as a POVM with $2^m$ distinct elements $\{P_{\tilde{X}}\}_{\tilde{X}=0}^{2^m-1}$, each corresponding to a unique choice of $\tilde{X}$. Thus the probability of Bob correctly determining whether a given state, encoding an input state chosen uniformly at random, corresponds to $X$ is 
\begin{eqnarray*}
P(\tilde{X}=X) &=& \frac{1}{2^m}\mbox{Tr}(P_{\tilde{X}} \rho_{\tilde{X}})+\frac{1}{2^m}\mbox{Tr}\left(\left(\mathbb{I} - P_{\tilde{X}}\right) \left(\mathbb{I}-\rho_{\tilde{X}}\right)\right)\\
&=& \frac{1}{2^m}\left(2^m -1 - \mbox{Tr}(P_{\tilde{X}} \left(\mathbb{I} - 2\rho_{\tilde{X}})\right)\right).
\end{eqnarray*}
If $e_{\tilde{X}}$ is the maximum eigenvalue of $\rho_{\tilde{X}}$ then the above probability is bounded from above by
\begin{eqnarray*}
P(\tilde{X} = X) &\leq& \frac{1}{2^{m}}\left(2^m -1- (1-2e_{\tilde{X}}) \mbox{Tr}\left(P_{\tilde{X}}\right)\right).
\end{eqnarray*}
However, as we have shown, for large $d$ the density matrix $\rho_{\tilde{X}}$ tends to a binomial distribution over $m+1$ states. Thus, the maximum eigenvalue of $\rho_{\tilde{X}}$ is given by $2^{-m} \binom{m}{\lfloor m/2 \rfloor}$ which approaches $\sqrt{2/\pi m}$. Therefore, for sufficiently large $m$ and $d$, we have 
\begin{eqnarray*}
P(\tilde{X} = X) &\leq& \frac{1}{2^{m}}\left(2^m -1- \left(1-\sqrt{\frac{8}{\pi m}}\right) \mbox{Tr}\left(P_{\tilde{X}}\right)\right).
\end{eqnarray*}
Averaged over all states this gives
\begin{eqnarray*}
\overline{P} &=& \frac{1}{2^m} \sum_{\tilde{X}=0}^{2^m - 1} P(\tilde{X}=X) \leq \sqrt{\frac{8}{\pi m}}.
\end{eqnarray*}
Thus the probability of Bob guessing Alice's input string is bounded from above by $\sqrt{8/\pi m}$ for sufficiently large $m$ and $d$.

The privacy of Bob's secret is more straight forward to prove. As Bob simply performs his secret operation upon Alice's input and returns it to her, the information Alice obtains is exactly the same as if she makes a single query to a black box function, and so Alice obtains the minimum possible information about Bob's secret unitary. The probability of Bob correctly determining Alice's input is substantially higher than the exponentially small bound one may hope for, but such a strong bound would violate the no-go theorems for oblivious transfer and bit commitment \cite{lo1997insecurity, spekkens2001degrees}. An alternate approach for Alice is to run many computations with different input states, where only one is her desired state and the remainder are dummies. However, this would allow Alice to extract more information about Bob's algorithm and is therefore less desirable for Bob.

\begin{figure}[!htb]
\includegraphics[width=0.92\columnwidth]{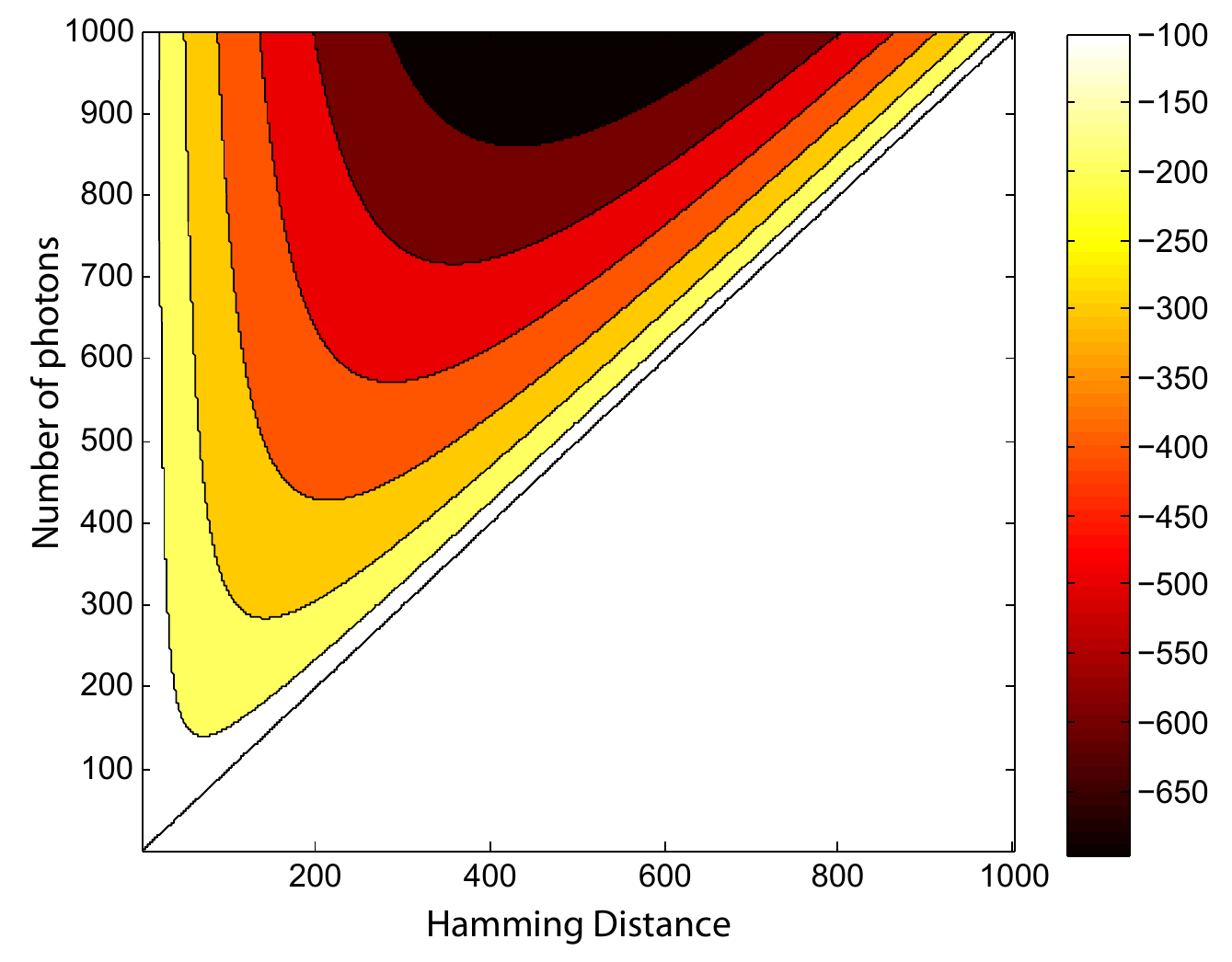}
\caption{(Colour online) $\mathrm{log}(\langle|\langle a| b\rangle|^2\rangle)$ with $d=1024$, against the number of photons $m$ and the Hamming distance between the strings.} \label{fig:overlap}
\end{figure}

\emph{The random attack ---} The average squared overlap between two states encoded with random keys is,
\begin{equation*}
\langle|\langle a|b\rangle|^2\rangle = \sum_\theta p(\theta) \mathrm{sin}^{2 h}(\theta) \mathrm{cos}^{2h'}(\theta),
\end{equation*}
where $h$ is the Hamming distance between strings $a$ and $b$, and $h+h'=m$. For a large number of divisions $d$, the overlap is plotted in Fig. \ref{fig:overlap}.

Note that the overlap is minimised when $h=m/2$. Thus it is easier to discriminate between states with Hamming distance close to $m/2$, and harder to distinguish states with lower or higher Hamming distance. One way Bob can make use of this property is to choose a key at random and measure all photons in this basis. As the measurement basis is virtually certain not to be unbiased with respect to the encoding basis, the string corresponding to the output of such a measurement will then be correlated with either the input string or its complement. Thus Bob can distinguish between states with Hamming distance sufficiently close to $m/2$.

Nonetheless Bob cannot perfectly infer Alice's secret input state if he has no prior information on the distribution. To see this, we note the overlap between $H$ or $V$, and a rotated $H$ or $V$, exhibits the property \mbox{$| \langle H | R(\theta) | H \rangle |^2 = | \langle V | R(\theta) | V \rangle |^2 = \mathrm{cos}^2\theta$}. Consequently, the probability of Bob's measurement results being perfectly correlated with Alice's secret state, given $m$ modes and $m$ photons, is \mbox{$| \langle \psi | R(\theta)^{\otimes m} | \psi \rangle |^2 = \mathrm{cos}^{2m}\theta$}, where $\ket\psi$ is Alice's input state and $\theta$ is that angle between Alice's chosen encoding basis and Bob's measurement basis.

\begin{figure}[!htb]
\includegraphics[width=0.75\columnwidth]{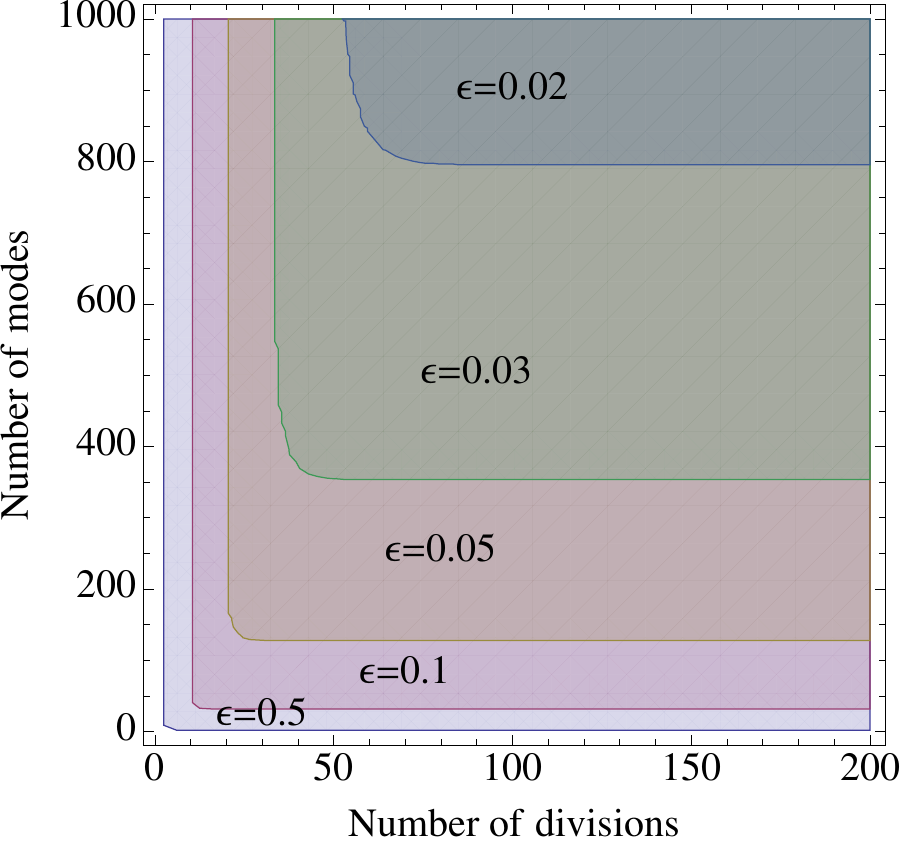}
\caption{(Colour online) Regions for different levels of confidence in the probability that Bob correctly infers Alice's state using a randomly chosen basis, $p_\mathrm{av}<\epsilon$.} \label{fig:region}
\end{figure}

If Bob choses a polarisation basis at random, the average probability that he will successfully infer the correct state is,
\begin{equation*}
p_\mathrm{av} = \frac{1}{d}\sum_{j=0}^{d-1} \mathrm{cos}^{2m}\left(\frac{j\pi}{d}\right).
\end{equation*}
Fig. \ref{fig:region} plots the value of this quantity for a range of values of $d$ and $m$. From it, two trends are clear. First, increasing $m$ decreases the probability of correctly identifying Alice's secret state. Second, increasing $d$ also decreases this probability, though it tends to a constant value, consistent with the bounds obtained from the Holevo information. For a large number of modes \mbox{$\lim_{m\to \infty} p_\mathrm{av} = 1/d$}, and for a large number of divisions \mbox{$\lim_{d\to\infty} p_\mathrm{av} = \Gamma(m+1/2)/\sqrt{\pi}m!$}, which scales as $1/\sqrt{\pi m}$ for large $m$. Thus this attack has a success probability close to the theoretical limit of $\sqrt{8/\pi m}$ .

\emph{Outlook \& conclusion ---} We note that the described approach to security is very specific to the Boson sampling and quantum walk models for LOQC, and will not work for the KLM protocol. This is because KLM requires adaptive measurement, which would require Alice disclosing the appropriate measurement basis to Bob in order for him to perform the appropriate measurement and feedforward. Thus, the security of this protocol relies on the unique property that there is \emph{no} measurement or feedforward \emph{within} the circuit.

A beneficial feature of this protocol is that only one round of communication is needed in each direction between Alice and Bob -- Alice prepares a mixed state, sends it to Bob to which he applies the computation and returns it back to Alice. This guarantees that the amount of information revealed about Bob's operation is no more than in the ideal case.

The described approach is technologically trivial. If we have the ability to implement Boson sampling or quantum walks, they can be encrypted simply with the addition of randomised wave-plate angles prior to and after the computation. Thus the ability to implement encryption of these protocols is foreseeable.

Our protocol relies on Alice performing random rotations about the \emph{y}-axis on the Bloch sphere. However it can be shown that more general rotations about a randomly chosen axis do not improve the asymptotic security of the scheme.

A key open question for the multi-walker quantum walk model is its applicability. Boson sampling can be regarded as an application of quantum walks. However, while shown to be likely classically hard to simulate, no specific algorithmic applications have been identified. While isomorphic to the Boson sampling model, the multi-walker quantum walk model may prove more fruitful for algorithm design, since it is inherently graph theoretic in nature and may therefore naturally lend itself to the development of graph theoretic algorithms.

We emphasise that our protocol does not guarantee that Bob learns nothing about Alice's data, but rather that the information Bob obtains is incomplete, asymptotically reducing Bob's probability of successfully reconstructing the input or output state. The trade-off that must be paid for improved security is a large number of randomised rotation settings and a larger interferometer.

In conclusion, we have presented a simple yet effective approach to encrypted quantum computation using two recent models for LOQC. The requirements for this protocol are well within current technological capabilities and could be readily implemented with present-day technology.


\emph{Acknowledgments ---} PR and AG acknowledge support from the Australian Research Council Centre of Excellence for Engineered Quantum Systems (Project number CE110001013). JF acknowledges support from the National Research Foundation and Ministry of Education, Singapore.

\bibliography{paper}

\end{document}